\documentclass[10pt,preprint]{aastex}

\def\gtrsim{\mathrel{\hbox{\rlap{\hbox{\lower4pt\hbox{$\sim$}}}\hbox{$>$}}}}
\def\lesssim{\mathrel{\hbox{\rlap{\hbox{\lower4pt\hbox{$\sim$}}}\hbox{$<$}}}}

\shortauthors{Tingay et al.}
\shorttitle{Brightness Temperatures of Extragalactic Radio Sources}

\begin{document}

\title{Measuring the Brightness Temperature Distribution of Extragalactic
Radio Sources with Space VLBI }

\author{S. J. Tingay\altaffilmark{1,2}, R. A. Preston\altaffilmark{2}, 
M. L. Lister\altaffilmark{2,3}, B. G. Piner\altaffilmark{2,4},
D. W. Murphy\altaffilmark{2}, D. L. Jones\altaffilmark{2},
D. L. Meier\altaffilmark{2}, T.J. Pearson\altaffilmark{5}, A. C. S. Readhead\altaffilmark{5}, H. Hirabayashi\altaffilmark{6}, Y. Murata\altaffilmark{6}, H. Kobayashi\altaffilmark{7}, and M. Inoue\altaffilmark{8}}

\altaffiltext{1}{Present address: Australia Telescope National Facility, P.O. Box 76,
Epping, NSW 2121, Australia}
\altaffiltext{2}{Jet Propulsion Laboratory, California Institute of Technology,
 MS 238-332, 4800 Oak Grove Drive, Pasadena, CA 91109-8099}
\altaffiltext{3}{Present address: National Radio Astronomy Observatory, 520 Edgemont Road, Charlottesville, VA 22903}
\altaffiltext{4}{Present address: Whittier College, Department of Physics, 13406
Philadelphia Street, Whittier, CA 90608}
\altaffiltext{5}{California Institute of Technology, 105-24, Pasadena, CA
91125}
\altaffiltext{6}{Institute of Space and Astronautical Science,
Yoshinodai 3-1-1, Sagamihara, Kanagawa 229-8510, Japan}
\altaffiltext{7}{National Astronomical Observatory, Ohsawa 2-21-1,
Mitaka, Tokyo 181-8588, Japan}
\altaffiltext{8}{Nobeyama Radio Observatory, Minamisaku, Nagano
384-1305, Japan}

\begin{abstract}

We have used VSOP space very long baseline interferometry observations
to measure the brightness temperature distribution of a well-defined
sub-set of the Pearson-Readhead sample of extragalactic radio sources.
VLBI which is restricted to Earth-diameter baselines is not generally
sensitive to emitting regions with brightness temperatures greater
than approximately $10^{12}$ K, coincidentally close to theoretical
estimates of brightness temperature limits, $10^{11} - 10^{12}$ K.  We
find that a significant proportion of our sample have brightness
temperatures greater than $10^{12}$ K; many have unresolved components
on the longest baselines, and some remain completely unresolved.
These observations begin to bridge the gap between the extended jets
seen with ground-based VLBI and the microarcsecond structures inferred
from intraday variability, evidenced here by the discovery of a
relationship between intraday variability and VSOP-measured brightness
temperature, likely due to the effects of relativistic beaming.  Also,
lower limits on jet Lorentz factors, estimated from space VLBI
observations, are starting to challenge numerical simulations that
predict low Lorentz factor jets.

\end{abstract}

\keywords{galaxies : jets ---
          galaxies : active ---
	  quasars : general ---
          radio galaxies : continuum ---
          radiation mechanisms : non-thermal ---
          techniques : interferometric}

\section{INTRODUCTION}

Ground-based very long baseline interferometry (VLBI) surveys
based on well-defined samples of extragalactic radio sources have been
of critical importance in determining the general morphological and
dynamical properties of the radio-bright jets originating in the
nuclear regions of active galaxies \citep{PR88,TVP94,MOE96,Kell98} and have
been essential in providing constraints and insights for theoretical
models of nuclear jets and their environments.  Studies of individual
sources alone may provide a biased view of nuclear activity.

The logical extension to this work is space-based VLBI surveys, which
allow higher resolution images to be achieved at a given observing
frequency.  Also, the ability to measure brightness temperatures
higher than possible from the ground, for a statistically useful
number of sources, is one of the major justifications for extending
the VLBI technique to space.  Brightness temperatures measured with
space VLBI can provide interesting constraints on jet formation models
and the theory of relativistic beaming, based on comparisons of
observed brightness temperatures with theoretical estimates of the
intrinsic brightness temperature upper limit.  If the jets in these
sources are relativistic and somewhat aligned with our line of sight,
the radiation from the jet can be apparently amplified, or Dopper
boosted, causing the measured brightness temperature to be apparently
in excess of the theoretical upper limit, yielding a lower limit on
the jet Doppler factor.

We identify three theoretical estimates of the intrinsic brightness
temperature upper limit, with which we will compare our observational
data in subsequent sections: the inverse Compton limit of \cite{KP69},
$10^{12}$ K; the equipartition brightness temperature of
\cite{Read94}, $10^{11}$ K; and the induced Compton scattering limit
of \cite{SK94}, $2\times10^{11}$ K.

The first attempt to observe radio-loud AGN using space VLBI
techniques was made by \cite{LEV86} using the TDRSS satellite in
conjunction with ground-based antennas. \cite{LIN89,LIN90} describe
measurements of brightness temperatures from these observations.  The
TDRSS observations demonstrated that space VLBI techniques were
feasible, but the limited number of ground-based antennas meant that
imaging was not possible.

We have conducted a space VLBI imaging survey using the facilities of
the VLBI Space Observatory Programme \citep[VSOP;][]{HH98,HH00},
drawing our sample from the well-studied Pearson-Readhead survey of
extragalactic radio sources \citep{PR81,PR88}.  Our survey has taken
advantage of long space VLBI baselines and large arrays of ground
antennas, such as the Very Long Baseline Array (VLBA) and European
VLBI Network (EVN), to achieve high resolution images of 27 Active
Galactic Nuclei (AGNs), and to measure the core brightness
temperatures of these sources more accurately than is possible from
the ground.  This work complements the VSOP full-sky survey, which
will observe most AGNs brighter than 1 Jy using a very limited set of
ground-based antennas
\citep{HS00}.

The Pearson-Readhead sample contains 65 northern radio sources
with 5 GHz flux densities greater than 1.3 Jy.  \cite{PR88} undertook
global 5 GHz VLBI observations of this complete sample, detecting 46
sources in total.  We have selected 31 of these on the basis of
long Earth baseline 5 GHz flux density, $>0.4$ Jy, as likely space
VLBI detections above the $7\sigma$ level between a VLBA antenna and the
orbiting antenna.

\section{OBSERVATIONS AND DATA ANALYSIS}

Observations using the HALCA orbiting antenna were conducted over the
period 1997 August to 1999 April at a frequency of 5 GHz, in
conjunction with either the VLBA, EVN, or in some cases, telescopes
from both arrays.  Typically HALCA was tracked between four and five
hours per observation, producing {\it (u,v)} coverages with a maximum
baseline of nearly 500 M$\lambda$, or $30,000$ km.  Full details of
these data, the fringe-fitting, calibration, and imaging
will be given in \cite{LIS00}. 

The final clean images and the corresponding self-calibrated datasets
were the basis for model-fitting analyses.  The set of point source
clean components describing each core were removed from the clean
component model and replaced with a single elliptical Gaussian,
described by 6 free parameters, which were fit to the {\it (u,v)}
plane data using the MODELFIT task in DIFMAP \citep{SPT94}.  The
visibility weights, derived from the scatter in the visibilities, used in this least squares fit were adjusted for
the space baselines so that the sum of the weights on the space
baselines equalled the sum of the weights on the ground baselines. In
this manner, the ground antennas and the space antenna contributed
equally to the reduced chi-squared statistic.  The adjustment 
of weights on the space baselines is essential.  Without adjustment
the high SNR ground visibilities would dominate the model-fit.  By
increasing the weights on the space baselines we ensure that we include
long baseline information in our Gaussian core models.  Runs of MODELFIT were
interspersed with phase self-calibration, to ensure that the reduced
chi-squared was minimized to find the best fit models.  

Errors on the parameters of the best-fit models were determined using
the DIFWRAP package \citep{lov00}.  We searched a four-dimensional
parameter space in component major axis, axial ratio, position angle,
and flux density for each source to determine which parameter ranges
fit the {\it (u,v)} data.  The sizes and flux densities of the
model-fitted core components, and the errors on these quantities, were
used to calculate limits on the observed brightness temperatures,
assuming a Gaussian surface brightness profile for the core.  We
convert the Gaussian brightness temperatures to optically thick sphere
brightness temperatures, by multiplying the Gaussian brightness
temperatures by a factor of $0.56$\footnote{The Gaussian and optically
thick sphere profiles are required to have the 50\% points of their
visibility functions occur at the same projected baseline, giving the
ratio between Gaussian FWHM and optically thick sphere diameter to be
1.6 \citep{TJP95}, resulting in the factor of 0.56 between the
corresponding brightness temperatures}.  The observed frame,
optically thick sphere brightness temperatures derived from our data
are tabulated elsewhere \citep{LIS00}.

To estimate lower limits on the Doppler factor we calculate
$\delta_{\rm{min}}\sim(T_{\rm{obs}}(1+z)/T_{\rm{int}})$, where
$\delta_{\rm{min}}$ is the minimum Doppler factor, $T_{\rm{obs}}$ is
the observed brightness temperature, $T_{\rm{int}}$ is
the intrinsic brightness temperature, and $z$ is the source redshift.
We discuss these lower limits in \S 3.1.  To estimate limits on the
brightness temperatures at a fixed frequency in the co-moving frame,
so that we can meaningfully compare all the sources, we calculate
$T_{\rm{co}}=T_{obs}(1+z)^{3-\alpha}$, where $\alpha$ is the
observed spectral index  ($S_{obs}\propto
\nu_{obs}^{\alpha}$) and $T_{\rm{co}}$
is the co-moving frame brightness temperature \citep{Read94}.  In the
discussion of the co-moving frame brightness temperatures in \S 3.2 we
adopt the optically thick case, $\alpha=2.5$, to correspond to the
physics embodied in the derivation of the theoretical brightness
temperature limits.  Support for considering the optically thick case
as a reasonable approximation for the observed sources comes from
those VSOP observations for which we have ground-based images at a
similar epoch and resolution but higher frequency, from which we
estimate core spectral indices.  For this small number of sources the
core spectral indices are substantially inverted.

Fig.~1 shows the resulting distribution of best fit brightness
temperatures in the co-moving frame, for optically thick sphere
profiles, and at a fixed frequency in the co-moving frame.  For many
sources the data provided only lower limits to the brightness
temperature.  The data for these sources are generally consistent with
a zero axial ratio (i.e., one-dimensional) component with position
angle equal to the position angle of the parsec-scale jet.  This is
possibly due to the presence of two or more closely spaced jet
components in the core region that cannot be resolved individually.
Data for 10\% of the sources in our sample ($0814+425$, $1624+416$,
and $1637+574$) are consistent with truly unresolved, point source
cores.  Although best-fit brightness temperatures were obtained with
these zero-axial ratio and point source fits, the data are consistent
within the errors with an infinite brightness temperature, giving only
a lower limit to the brightness temperature.  The set of flux vs {\it
(u,v)} distance plots of our data given in \cite{PRE00} illustrate
this result in a different way, showing that the jets of many sources
are resolved out on baselines just longer than an Earth diameter,
leaving only a marginally resolved or unresolved core component on the
longest space baselines.  For two sources, 4C 39.25 and 2021$+$614, the cores were too weak to constrain the model-fitting and both have unbounded upper and lower brightness temperature limits.

\section{DISCUSSION}

\subsection{Upper limits to brightness temperature and lower limits to
the Doppler factor} 

The highest brightness temperature lower limit we have found is
$1.8\times10^{12}$ K, for 0133$+$476 (in the observer's frame),
prompting us to explore the physical mechanisms which can produce
brightness temperatures in excess of theoretical intrinsic limits.
The most popular current model is Doppler beaming, as discussed in the
introduction.  In this case our estimates of flux density are
exaggerated, leading us to infer brightness temperatures higher than
in reality.  It is possible for a typical jet Lorentz factor
$\Gamma=[1-(v/c)^2]^{-1/2} = 5$ to boost the observed brightness
temperatures up by a factor of $2\Gamma = 10$.

\cite{KP69} derived an upper limit to the intrinsic brightness temperature of
approximately $10^{12}$ K, based on the inverse Compton losses at high
photon energies that occur when a synchroton plasma approaches
$10^{12}$ K.  The brightness temperature lower limits of three of our
sources lie above $10^{12}$ K, implying the existence of Doppler
boosting even in this extreme theoretical limit.

More recent theoretical work has shown that rather extreme conditions
must be present in order for a synchrotron plasma to have an intrinsic
brightness temperature near the inverse Compton limit. \cite{Read94}
and \cite{BRS94} point out that the ratio of energy density in
relativistic electrons to the energy density of the magnetic field
must be extremely large (typically $\sim 10^7$) for the inverse
Compton catastrophe to occur.  For sources with milder energy density
ratios (including those at equipartition, where $u_{rel} \simeq
u_{B}$), \cite{Read94} has shown that their intrinsic brightness
temperatures should lie close to  $10^{11}$ K, which is an order of
magnitude less than the inverse Compton limit. \cite{SK94} have also
considered the effects of induced Compton scattering by relativistic
electrons in an AGN jet, and derive a hard limit to the brightness
temperature of a self-absorbed synchrotron source to be approximately
$2 \times 10^{11}$ K. Although one cannot completely rule out the
possibility that some AGNs may have extremely large $u_{rel}$ to
$u_{B}$ ratios, it seems unlikely that this scenario is typical of the
general AGN population, given the short lifetimes of sources radiating
at the inverse Compton limit.

For 0133$+$476, the inverse Compton scenario gives a lower limit to
the Doppler factor of approximately $3$ and an upper limit of $\sim
19\arcdeg$ on the jet viewing angle.  If we adopt a more likely
intrinsic brightness temperature of $10^{11}$ K, our estimate of the
Doppler factor lower limit increases to $\sim 30$.  The jet would have
to be aligned to within only $2\arcdeg$ of the line of sight in this
case and the Lorentz factor lower limit would be $\Gamma>15$.
As numerical modeling of relativistic jets becomes more
sophisticated, it appears to be difficult to produce jets from the
vicinity of black holes with $\Gamma \gtrsim 3$
\citep{KMSK00}, which corresponds to the escape velocity from the inner
edge of an accretion disk.  These values are also much lower than the
$\Gamma$'s of up to $\sim 40$ needed to explain superluminal motions
\citep[e.g.,][]{MMM00}. High Lorentz factor jets require the rapid 
acceleration of very light material outside the disk, such as a 
corona \citep{MEGPL97} or particles created near the horizon \citep{BZ77}. 
Even these processes are limited to $\Gamma \sim 10$ or so near 
the black hole by the drag force of photons emitted from the accretion 
disk \citep{LB95}, although continued acceleration to higher speeds 
well beyond the central black hole is still possible (see, {\it e.g.}, 
\citealt*{HB00}).  Therefore, direct measurements of brightness
temperatures with space VLBI are beginning to challenge some
theoretical models for relativistic jet formation in the inner regions
of AGNs.

Alternatively, if non-equilibrium conditions exist then it is
theoretically possible to exceed $10^{12}$ K for a synchrotron plasma
\citep{Sly92}.  To exceed this value for long periods of time, energy
would need to be continuously injected into the base of the jet.
Other suggested mechanisms for producing high brightness temperatures
in AGNs include coherent emission \citep{BL98} and conical shocks
\citep{SSP99}.

\subsection{Correlations with other source properties}
We have gathered a vast amount of supporting data from the literature
on our sample, in order to look for possible correlations between
VSOP-measured quantities and other source properties.  A full
multi-dimensional correlation analysis of our sample will be presented
in \cite{LTP01}. Here we discuss our results concerning the VSOP
core brightness temperatures.

As many of our brightness temperature measurements involve lower
limits, we used a non-parametric Kendall's tau test from the ASURV Rev
1.2 survival analysis package \citep{ASURV} to establish the
statistical likelihood of possible correlations.  Interestingly, we
did not find any correlations between co-moving frame brightness
temperature and potential beaming indicators such as core dominance,
jet bending, spectral index, emission line equivalent width, and
percentage optical polarization. This may be due to the large number
of lower limits in our dataset, which hinders the detection of
statistically robust correlations.

We have also performed a series of Gehan's generalized Wilcoxon
two-sample tests \citep{G65} on our data, to investigate whether
objects of various AGN classes have different co-moving frame
brightness temperatures, at the 95\% confidence level. We find no
statistically significant differences in the distributions of a)
quasars versus BL Lac objects, b) high-optically polarized quasars
($m_{opt} > 3\%$) versus low-optically polarized quasars, and c)
EGRET-detected versus non-EGRET detected sources.

One source property which does appear related to co-moving frame
brightness temperature is intraday variability (IDV). This phenomenon,
in which a source displays variability on timescales of a day or less,
occurs in roughly $30\%$ of all compact radio sources \citep{Q92}.
The rapid nature of these fluctuations implies very small source
sizes, and in turn, high brightness temperatures \citep{DEN00,KED97}. \cite{Q92} have divided IDV
activity into three classes: class II consists of rapid up-and-down
flux density fluctuations, class I involves monotonic increases or
decreases on short timescales, and class 0 indicates no IDV.  In
\cite{LIS00} we list the IDV classification for the sources in
our sample that have been monitored for IDV.  Some sources display
different classes of IDV activity at different times: for these
objects we list the maximum class of IDV published in
\cite{Q92,Q00} and \cite{KQW92}.  We find that the class II sources
have higher co-moving frame core brightness temperatures than the
class 0 and class I sources, at the $98.0\%$ and $97.8\%$ confidence
levels, respectively. 

Although this finding appears to imply a close relationship between
the compactness of the core and IDV activity, the co-moving frame
brightness temperatures are significantly smaller than those implied
from the IDV timescales (i.e., $10^{15-21}$ K), if they are intrinsic
to the source.  Part of this difference might be due to the differing
effects of Doppler factor on the brightness temperature estimates.
Brightness temperature estimates from IDV should be a factor of
$\delta^{2}$ higher than from VLBI, since they depend on timescale
measurements \citep{V94}.  Any remaining difference may be explained
if a contribution to the observed IDV is due to interstellar
scintillation \citep{KED97}.


\cite{Q92} find that virtually all sources with compact VLBI
structures show IDV at a level greater than $2\%$ of total flux,
indicating a relationship between IDV and relativistic beaming.  We
have found further evidence for such a relationship by using space
VLBI to probe these compact structures, measuring core brightness
temperatures, which are directly related to the Doppler factors of
their jets.  It seems likely that the relationship we find between IDV
type and VSOP-measured brightness temperature is partly due to Doppler
boosting.  High Doppler factors will boost brightness temperatures to
high values and will also decrease the timescale of any intrinsic
variations in flux, as seen in the observer's frame; a timescale
intrinsic to the source of $\Delta t$ corresponds to a timescale in
the observer's frame of $(1+z)\Delta t/\delta$.  This could tend to
give highly beamed sources high apparent brightness temperatures and
type II IDV, whereas less beamed sources will have lower apparent
brightness temperatures and type 0 or I IDV.
  
We gratefully acknowledge the VSOP Project, which is led by the
Japanese Institute of Space and Astronautical Science in cooperation
with many organizations and radio telescopes around the world.  Part
of this work was undertaken at the Jet Propulsion Laboratory,
California Institute of Technology, under contract with the National
Aeronautics and Space Administration.  SJT acknowledges support
through an NRC/NASA-JPL Research Associateship.  Thanks go to Stefan 
Wagner for providing updated IDV information for 1823$+$568.

\vfill\eject

\begin{figure}

\epsscale{0.85}
\plotone{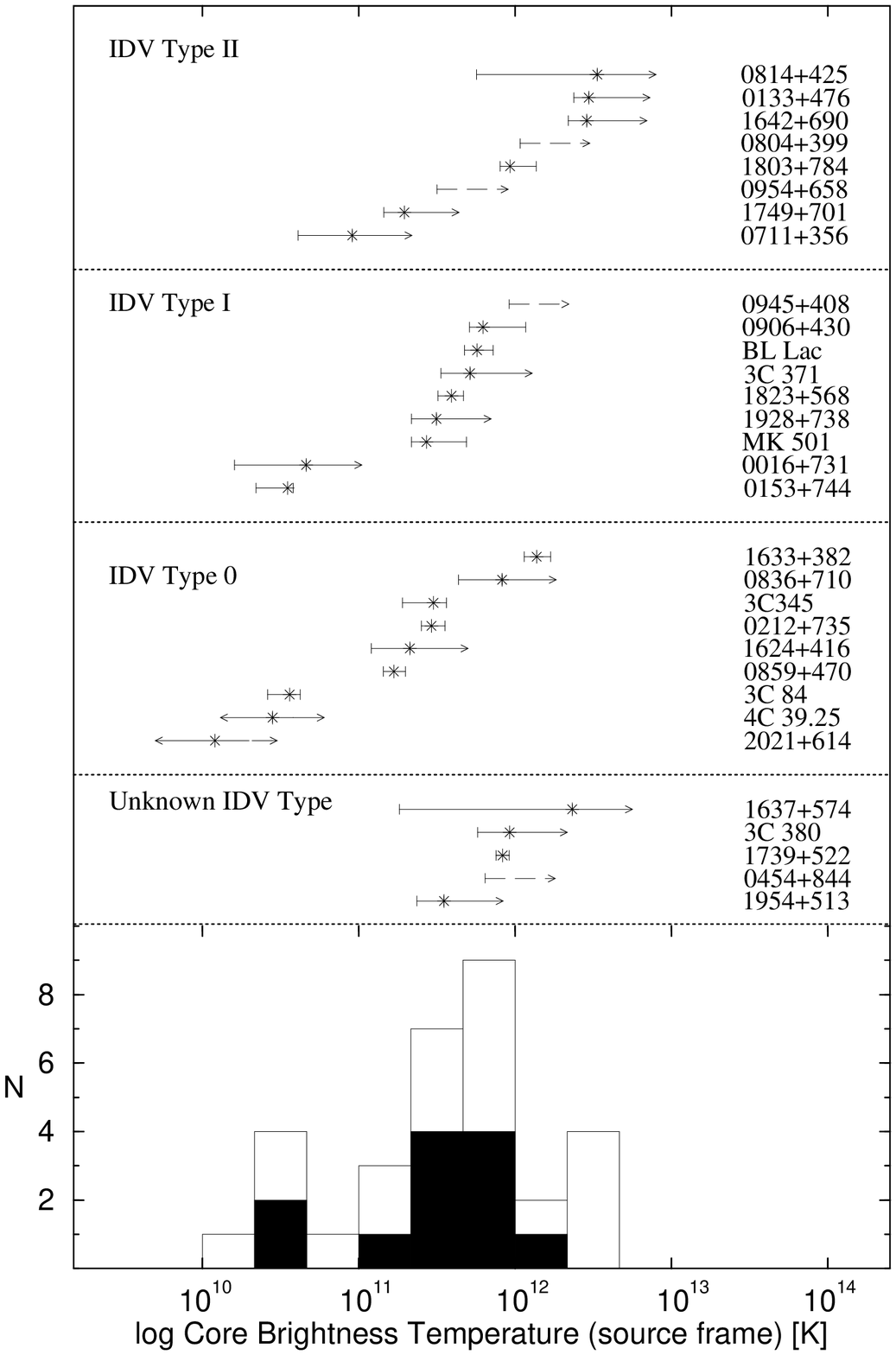}

\caption{Top panels: plots of co-moving frame, optically thick brightness temperatures for the cores of Pearson-Readhead
AGNs, grouped by IDV class. Sources indicated with dashed
lines have not yet been observed with VSOP. Bottom panel: histograms
of best-fit core brightness temperature for the entire sample
(unshaded), and only those sources with measured lower and upper
limits on brightness temperature (shaded).}
\end{figure}

\end{document}